\documentclass[journal]{IEEEtran}
\usepackage{bm}
\usepackage{theorem}
\usepackage{enumerate}
\usepackage{multirow}
\usepackage{kbordermatrix}
\usepackage{caption}
\usepackage{psfrag}
\usepackage{float}
\usepackage{algorithm}
\usepackage[noend]{algpseudocode}

\newcommand{\bea}[1]{\begin{eqnarray} #1 \end{eqnarray}}
\newcommand {\eqrefn}{Eq.~\eqref}

 \newcommand{\ebf}{\mathbf{e}}

 \newcommand{\zbf}{\mathbf{z}}

\newcommand{\Var}{\mathrm{Var}}

\newcommand{\mubf}{\boldsymbol{\mu}}
\newcommand{\Sigmabf}{\boldsymbol{\Sigma}}\newcommand{\epsilonbf}{\boldsymbol{\epsilon}}
\newcommand{\lambdabf}{\boldsymbol{\lambda}}
\newcommand{\deltabf}{\boldsymbol{\delta}}

\newcommand{\Abf}{\mathbf{A}}  \newcommand{\Cbf}{\mathbf{C}}
 \newcommand{\Ebf}{\mathbf{E}} 
  \newcommand{\Ibf}{\mathbf{I}}
  \newcommand{\Lbf}{\mathbf{L}}
  
 \newcommand{\Qbf}{\mathbf{Q}} \newcommand{\Rbf}{\mathbf{R}}
\newcommand{\Sbf}{\mathbf{S}}  \newcommand{\Ubf}{\mathbf{U}}
\newcommand{\Vbf}{\mathbf{V}}  
 \newcommand{\Zbf}{\mathbf{Z}}

\newcommand{\zeros}{\textbf{0}}

{\theoremstyle{break}\theoremheaderfont{\normalfont\bfseries}}
{\theoremstyle{plain}\theorembodyfont{\normalfont\rmfamily}\newtheorem{pf}{Proof}}
{\theoremstyle{break}\theoremheaderfont{\normalfont\bfseries}
{\theoremstyle{break}\newtheorem{proposition}{Proposition}\theoremheaderfont{\normalfont\bfseries}
{\theoremstyle{break}\theoremheaderfont{\normalfont\bfseries}
{\theoremstyle{break}\theoremheaderfont{\normalfont\bfseries}
{\theoremstyle{break}\theoremheaderfont{\normalfont\bfseries}

\usepackage{cite}

\ifCLASSINFOpdf
\usepackage[pdftex]{graphicx}
\else
\usepackage[dvips]{graphicx}
\fi
\usepackage{amsmath}

\usepackage{array}

\usepackage{url}

\hyphenation{op-tical net-works semi-conduc-tor}

\begin{document}
%
\title{Identifying Topology of Power Distribution Networks Based on Smart Meter Data}
%
%
%

\author{Satya~Jayadev~P,~\IEEEmembership{Student Member,~IEEE},~Nirav~Bhatt,~\IEEEmembership{Member,~IEEE},\\~Ramkrishna~Pasumarthy,~\IEEEmembership{Member,~IEEE}~and~Aravind~Rajeswaran 
        
\thanks{The finance support to Satya Jayadev P. and Aravind Rajeswaran from Data Science Initiative Grant of IIT Madras, and Nirav Bhatt from Department of Science \& Technology, India through INSPIRE Faculty Fellowship is acknowledged.}%
\thanks{ Satya Jayadev P and Ramkrishna Pasumarthy are with Department of Electrical Engineering, Nirav Bhatt is with the Department of Chemical Engineering, Indian Institute of Technology Madras, India and Aravind Rajeswaran is with ILDS group of Department of Computer Science \& Engineering. ee15d202@smail.iitm.ac.in, niravbhatt@iitm.ac.in, ramkrishna@ee.iitm.ac.in, aravindr@smail.iitm.ac.in,  }}%

\maketitle
\begin{abstract}
In a power distribution network, the network topology information is essential for an efficient operation of the network. This information of network connectivity is not accurately available,  at the low voltage level, due to uninformed changes that happen from time to time. In this paper, we propose a novel data--driven approach to identify the underlying network topology including the load phase connectivity from time series of energy measurements. The proposed method involves the application of Principal Component Analysis (PCA) and its graph-theoretic interpretation to infer the topology from smart meter energy measurements. The method is demonstrated through simulation on randomly generated networks and also on  IEEE recognized Roy Billinton distribution test system. 
\end{abstract}


%
\IEEEpeerreviewmaketitle
\section{Introduction}
\IEEEPARstart{T}{he} complexity of power distribution networks is increasing day by day with advancements in technology and addition of new and sophisticated components to the power grid. The concentration of power systems control and monitoring has traditionally been at the generation, transmission and high voltage distribution levels. A need for advancement in control at the low voltage (LV) distribution level arose with the advent of active distribution networks with intermittent distributed resources such as solar, wind energy etc., and plug-in devices such as electric vehicles. Research is  actively pursued in the areas of control and automation of distribution networks, and in many cases, it is assumed that network topology information is available \cite{Schneider, Jiyuan09, Kersting07}.

The topology of an LV distribution network gives the connectivity among its numerous assets such as feeders, distribution transformers, distributors and consumers. The information of the underlying network topology is useful for  efficient integration of renewable energy sources  and efficient  management of outages in  distribution networks \cite{Lueken12,Melo13}. Further, for a reliable state estimation in a distribution network, accurate information of the network topology is essential \cite{Cavaro15}.

Some of the LV consumers operate on single-phase and they draw power from one of the three phases of a distribution transformer. The phase connectivity of those consumers also forms a part of the network topology information. This information is important in maintaining load and voltage balances in the three phases of the distribution transformers and the distribution feeders. Unbalanced loads on transformers and feeders lead to higher copper losses and voltage drop, and consequently affect the life of the assets \cite{Dickson09}.

The network topology information might not be accurately available at all time because of changes that take place due to network reconfiguration, repairs, maintenance and load balancing \cite{Das06, Dickson09}. Moreover, the consumers might have a facility to switch between phases when a phase trips, and, thus changing the topology. Often, the network operators are not aware of such changes in the topology \cite{Jing12}. 

A number of attempts were made to solve the problems of phase identification, and topology identification. Smart grid technologies have further intensified the search for new methods of inferring connectivity. 
The methods for topology identification can be classified into two categories: (i) Hardware based methods, and (ii) software based methods.

Hardware based methods include microprocessor based phase identification system and signal injection device designed for phase measurement \cite{Chen11,Zhiyu13}. However, the additional hardware and staff required for these devices to work, makes these options costly. 

Software based methods have become popular with the advent of Advanced Metering Infrastructure (AMI) such as smart meters and Phasor Measurement Units (PMU). These devices are installed at important nodal points and they generate large amount of data at regular time intervals which can be collected and analysed at centralised data centres. In the literature, researchers have proposed methods for analysing these data for topology identification. 

There were some software based methods which were presented prior to the development of AMI. One is a search algorithm to determine phase information using power flow measurements and load data \cite{Dilek02}. Its drawback is that it ignores noise and uncertainty in data. Another method  proposed real--time monitoring of changes in the underlying network topology based on the status of circuit breakers\cite{Kezunovic06}.  

The latest methods include optimization based approach to infer phase connectivity and also network topology from time series of power measurements \cite{Arya11, Arya13}. The authors proposed Mixed Integer Programming (MIP) based solution which is computationally intensive to solve without relaxing the constraints. In \cite{Pezeshki12}, a technique to identify the phases based on cross-correlation method using the time series of voltage measurements, is presented. \cite{Tom13} proposed a linear regression based algorithm for phase identification which considers the correlation between consumer voltage and substation voltage. It requires the Geographical Information System (GIS) model which may not always be available. In another method for phase identification, data obtained from micro synchrophasor measurement units ($\mu$PMU) is analysed \cite{Wen15}. In \cite{Cavaro15,Wiel14}, methods were proposed to infer topology from time series of measurements from PMUs. In \cite{Bolognani13}, an algorithm to identify the network topology using voltage correlation analysis is presented. In \cite{erseghe2013topology}, a hypothesis testing based technique to the topology identification is proposed using the signals generated by the PLC network laid in a smart grid. 

In this work, we propose a novel method for identifying the underlying network topology from smart meter energy measurements. Our approach integrates PCA and its graph-theoretic interpretation (as shown in \cite{Aravind15}) to identify the network topology from time series of energy measurements, based on the principle of energy conservation. As a preliminary, in our prior work \cite{Jayadev16}, we introduced this approach for phase identification. In this paper, we elucidate our method in detail and then extend it to solve a problem of  network topology identification. 

The rest of the paper is organised as follows. Section \ref{Prelims} revisits some necessary preliminaries. The mathematical formulation of the problem is given in Section \ref{ProbForm}. The proposed solution and algorithms are presented in Section \ref{Solution}. Finally, the simulation results and conclusions are provided in Sections \ref{Simulation} and \ref{Conclusion}, respectively.

\section{Preliminaries} \label{Prelims}
\subsection{Principal Component Analysis (PCA)}
PCA is one of the widely used tools of multivariate data analysis with many applications. It has been primarily developed as a method for denoising data and 
dimensionality reduction \cite{Jolliffe02}. Recently, PCA has been applied to identify a model in the presence of noise\cite{Jolliffe02, Narasimhan08}.  In this paper, we will use the method of model identification using PCA from measurements and, we will revisit it next. 
\subsubsection{Model Identification using PCA} \label{ModelID}
Let  $\zbf^m(j)$ be defined as a sample of $n$ variables measured at the $j^{th}$ time instance, as follows:
\begin{eqnarray}
\zbf^m(j) &=& \left[\begin{array}{c} z_1^m(j), \, z_2^m(j), \,\cdots, z_n^m(j) \end{array}\right]^T
\end{eqnarray}
Generally, the measured values are corrupted by random noise leading to an error in the samples. The vector of measured variables can be written as:
\begin{equation}
\zbf^m(j) = \zbf^t(j) + \ebf(j)
\end{equation}
where $\zbf^t(j)$ is the vector of true values of the variables at the $j^{th}$ time instance and $\ebf(j)$ is the vector of errors due to noise. It is assumed that the error is normally independent and identically distributed (\emph{i.i.d.}) as follows: 
\begin{equation}
\ebf(j) \sim \mathcal{N}(\zeros,\sigma^2_e \Ibf) \label{Eq:errordist}
\end{equation}
where $\sigma_e^2$ is an error variance, $\mathcal{N}$ indicates the Gaussian distribution, and $\Ibf$ is the $(n\times n)$ identity matrix.
The $n$ variables are linearly related  by the following model: 
\begin{equation}
\Cbf \, \zbf^t(j) = \zeros \label{Eq:ConstrainedModel}\vspace{-0.1cm}
\end{equation}
where $\Cbf$ is a $(p \times n)$-dimensional  constraint matrix, with $p$ being the number of linear relationships. 
From measurement vectors available at $N$ time instants, an $(n \times N)$-dimensional $\Zbf$ matrix can be constructed by stacking $\zbf^m(j),\,\,j=1,2,\ldots,N$ vectors. \eqrefn{Eq:ConstrainedModel} indicates that the noise-free data lies in an $(n-p)$-dimensional subspace orthogonal to the $p$-dimensional subspace spanned by the  rows of the $\Cbf$. 
The objective of model identification using PCA is to estimate the $(n-p)$-dimensional true data subspace and the $p$-dimensional constrained subspace, given the data matrix $\Zbf$.

In PCA, these subspaces are obtained from the eigenvectors of the covariance matrix  $\Sbf_z = \Zbf\Zbf^T$. The subspaces are identified such that the sum of the squared difference between the measured values and denoised estimates of the values of the variables is minimised \cite{Narasimhan08}. 
The eigenvectors of the covariance matrix can be determined using Singular Value Decomposition (SVD) of $\Zbf$ as follows:
\begin{equation}
\text{SVD}(\Zbf) = \Ubf_1\Sbf_1\Vbf_1^T + \Ubf_2\Sbf_2\Vbf_2^T \label{Eq3}
\end{equation}
where $\Ubf_1$ is the set of orthonormal eigenvectors corresponding to the $(n-p)$ largest eigenvalues of $\Sbf_z$ while $\Ubf_2$ is the orthogonal eigenvectors corresponding to the smallest $p$ eigenvalues of $\Sbf_z$. $\Sbf_1$ and $\Sbf_2$ are diagonal matrices with the singular values of $\Zbf$. It has been shown that $\mathcal S_R(\Ubf_2^T) \sim \mathcal S_R(\Cbf)$, where $\mathcal S_R(.)$ indicates the subspace spanned by the rows of $(.)$ matrix \cite{Narasimhan15}. Then, $\Ubf_2^T$ satisfies the following relationship:
\begin{equation}
\Ubf_2^T\zbf=0 
\end{equation}
Hence, $\Ubf_2^T$ gives the constraint matrix and it is to be observed that the constraint matrix suffers from the rotational ambiguity: 
\begin{equation} \label{estimate}
\Ubf_2^T\zbf = \hat{\Cbf}\zbf = \Qbf\hat{\Cbf}\zbf=\zeros
\end{equation}
where $\Qbf$ is a non-singular matrix. Hence, the estimated constraint matrix $\hat{\Cbf}$ is not unique and may not be the one which has direct physical interpretation.

A regression model can also be obtained by partitioning the variables into  a set of dependent variables $\zbf_d$ having dimension ($n_d=p$) and independent variables $\zbf_i$ ($n_i=n-p$). The columns of $\hat{\Cbf}$ corresponding to the $\zbf_d$ and $\zbf_i$ can also be partitioned as follows: $\hat{\Cbf} =[\hat{\Cbf}_d, \hat{\Cbf}_i]$, where  $\hat{\Cbf}_d$ and $\hat{\Cbf}_i$ are the $(n_d\times n_d)-$ and $(n_d \times n_i)-$dimensional matrices, respectively. Then, from \eqrefn{estimate} we obtain
\bea{\hat{\Cbf}_d\zbf_d +\hat{\Cbf}_i\zbf_i=\zeros. \label{DepInd}}
Since $\Ubf_{2d}$ is of full rank, we can express \eqrefn{DepInd} in terms of the estimated regression matrix relating the dependent and independent variables as follows:
\bea{\zbf_d =-(\hat{\Cbf}_d)^{-1}\hat{\Cbf}_i\zbf_i=\hat{\Rbf}\zbf_i, \label{regression}}
where $\hat{\Rbf}$ is the $(n_d \times n_i)$--dimensional regression matrix. The regression matrix $\hat{\Rbf}$ is proven to be unique \cite{Narasimhan15}.

The estimate of subspace of $\Cbf$ is not optimal in maximum likelihood sense when the assumption of \emph{i.i.d.} error in \eqrefn{Eq:errordist} does not hold, $\ebf(j) \sim \mathcal{N}(\zeros,\Sigmabf_e)$. In such cases, two approaches proposed in \cite{Narasimhan08}  can be used to estimate  $\Cbf$. Next, we will describe briefly one of the approaches when the error covariance matrix $\Sigmabf_e$ is known. For details of both approaches, refer \cite{Narasimhan08}. The approach transforms the data matrix by scaling it with cholesky factor of the error covariance matrix. Cholesky decomposition of  $\Sigmabf_e$ is given by:
\begin{equation} \label{Cholesky}
\Sigmabf_e = \Lbf\Lbf^T \vspace{-0.2cm}
\end{equation}
where $\Lbf$ is the $(n \times n)$--dimensional lower triangular matrix. The noisy data matrix is transformed into $\Zbf_s$ as follows:
\begin{equation}
\Zbf_s = \Lbf^{-1}\Zbf = \Lbf^{-1}\Zbf_t + \Lbf^{-1}\Ebf
\end{equation}
where $\Ebf$ is the error matrix, and $\Zbf_t$ is the data matrix having the true values. The covariance matrix of the transformed  matrix is:
\begin{equation}
\Sbf_{zs} = \Zbf_s\Zbf_s^T
\end{equation}
It is shown that by applying PCA on $\Zbf_s$, we get an estimate of the constraint matrix $\Cbf$ pertaining to the transformed data, on which inverse transformation is applied to get constraint matrix corresponding to original data \cite{Narasimhan15}. We apply PCA to $\Zbf_s$ and get an estimate of the constraint matrix and the regression matrix corresponding to the original data as follows:
\begin{eqnarray}
\text{SVD}(\Zbf_s) &=& \Ubf_{1s}\Sbf_{1s}\Vbf_{1s}^T + \Ubf_{2s}\Sbf_{2s}\Vbf_{2s}^T \\
\hat{\Cbf} &=& \Ubf_{2s}^T \Lbf^{-1} \\	\label{Constraints}
\hat{\Rbf} &=& -(\hat{\Cbf}_d)^{-1}\hat{\Cbf}_i \vspace{-0.85cm}
\end{eqnarray}
\subsection{Graph Theory Overview}
In this section, certain concepts of algebraic graph theory pertaining to this work are revised. 
\subsubsection{Graph and Sub-Graph}
A graph $G=(N_G,\,N_E)$ contains a set of nodes ($N_G$) and edges ($N_E$), whose connectivity represents a network of physical or abstract elements. A graph is said to be directed if its edges are directed from one node to another. $S$ is a sub-graph of $G$ with set of nodes $N_S \subset N_G$ and set of edges $E_S \subset E_G$ such that $E_S$ contains all edges with both end points in $N_S$.

A graph is said to be connected if there exists a path between every pair of its nodes, otherwise the graph is disconnected. The connected sub-graphs of a disconnected graph are referred to as its components. 
\subsubsection{Tree}
\begin{figure}[h]
	\centering
	\psfrag{P}[c][][2]{\begin{tabular}{@{}l@{}}
			Parent\\
		Nodes	
		\end{tabular}}
		\psfrag{C}[c][][2]{Child Nodes}
		\psfrag{P1}[c][][2]{P1}
		\psfrag{P2}[c][][2]{P2}
		\psfrag{P3}[c][][2]{P3}
		\psfrag{C1}[c][][1.5]{C1}
		\psfrag{C2}[c][][1.5]{C2}
		\psfrag{C3}[c][][1.5]{C3}
		\psfrag{C4}[c][][1.5]{C4}
		\psfrag{C5}[c][][1.5]{C5}
		\psfrag{C6}[c][][1.5]{C6}
		\psfrag{C7}[c][][1.5]{C7}
		\psfrag{C8}[c][][1.5]{C8}
		\psfrag{C9}[c][][1.5]{C9}
		\psfrag{a}[c][][2]{a}
		\psfrag{b}[c][][2]{\hspace{1mm}b}
		\psfrag{c}[c][][2]{c}
		\psfrag{d}[c][][2]{d}
		\psfrag{e}[c][][2]{e}
		\psfrag{f}[c][][2]{f}
		\psfrag{g}[c][][2]{g}
		\psfrag{h}[c][][2]{h}
		\psfrag{i}[c][][2]{i}
		\includegraphics[scale=0.5,width=80mm]{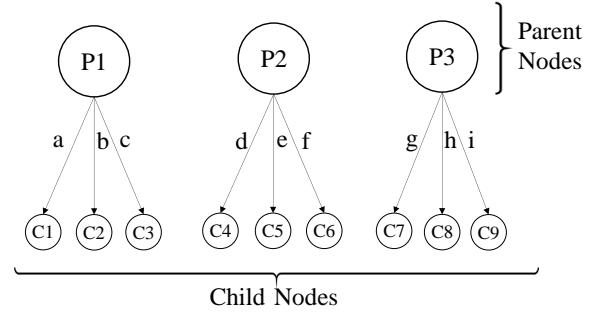}
\caption{A Forest with three tree components}\label{Figure0}	
	\end{figure}

A tree is a graph with no circuits (or loops) and any two nodes in a tree can be connected by a unique path. A directed tree is similar to a directed graph with its edges directed from one node to another. A disconnected graph with its components as trees is called a forest. Fig.~\ref{Figure0} shows a directed forest with three tree components. 

In a directed graph or tree, a parent node is a node which has an edge directed to another node(s) called the child node(s). In Fig.~\ref{Figure0}, the nodes 4, 5 and 6 are child nodes to the parent node 1. 

%

\subsubsection{Incidence Matrix}
The incidence matrix ($\Abf$) of a graph, $G=(N_G,\,N_E)$, describes the incidence of edges on nodes and is defined as follows for a directed graph:
\begin{equation}
\Abf_{n \times e} =
\begin{cases}
+1, \; \text{if edge} \; j \; \text{enters node} \; i \\ 
-1, \; \text{if edge} \; j \; \text{leaves node} \; i \\
0, \; \text{if edge} \; j \; \text{is not incident on} \; i
\end{cases}
\end{equation}
where $n = |N_G|$ and $e = |N_E|$

\begin{proposition} \label{Prop}
A directed graph (or a directed forest) can be uniquely constructed from an incidence matrix, provided there are no self loops. 
\end{proposition}

\begin{pf}
The proof of the Proposition is similar to the proof of Theorem 8 in Chapter 3 in \cite{Andrasfai91}.
\end{pf}

\subsection{Types of Distribution Networks} \label{TypesDistNet}
Based on the topology of the network, the distribution networks are classified as:
\begin{enumerate}[i]
\item {\bf Radial distribution network:}  In this configuration, each of the feeders and distributors is fed by a single source. Hence, this type of network is characterized by existence of a unique path from the source (substation) to each of the consumers.
\item {\bf Ring main distribution network:} In this configuration, the feeders and distributors may be connected to multiple sources for higher reliability of power supply. Hence, multiple sources are available for feeding a load, and there may be multiple paths between such sources and loads. However, during network operation, the circuit breakers are configured such that only one source feeds a load, and an electrically active path between them is unique. Hence, the active network can still be considered to be \emph{radial}. 
\end{enumerate}

\section{Problem Formulation} \label{ProbForm}
\subsection{Distribution Network as a Tree} \label{DistNw}

\begin{figure}[h]
\centering
\psfrag{S}[c][][1.75]{\tiny Substation Node}
\psfrag{F}[c][][1.75]{\begin{tabular}{@{}l@{}}
		\\
  \tiny Feeder \& \\
 \tiny  Transformer \\
 \tiny  Nodes
\end{tabular}}{ }
\psfrag{C}[c][][1.75]{\tiny Consumer Nodes}
\includegraphics[scale=0.25]{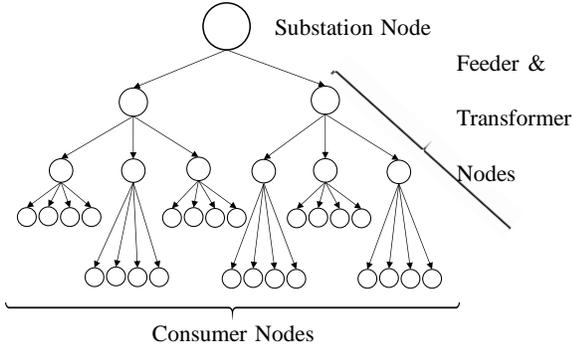}
\caption{Tree representation of Network Topology}\label{Figure1}
\end{figure}

The topology of a distribution network can be considered to be the connectivity  between the meters installed at the substation, feeders, transformers, and consumer mains. A graph can be constructed by assigning nodes to each of the meters and the connections between them can be represented as edges. Since the paths from the substation  to each of the consumers are unique as described in Section \ref{TypesDistNet}, the graph of a distribution network is a tree as shown in Fig.~\ref{Figure1}. For the meters of 3-phase loads and 3-phase transformers, we  assign three separate nodes corresponding to each of the three phases, in order to maintain consistency with the single-phase loads. The main advantage of such an assignment is that it also allows us to determine the phase identity of single-phase consumers\footnote{In this paper, the words `meters', `nodes' and `variables' are interchangeable. Also the words `samples' and `readings' are interchangeable.}. 

\subsection{Energy measurements}
Since we assume that smart meters are installed at all the nodal points in the network,  energy measurements in watt-hour (Wh) are obtained from the meters over for regular time intervals, generally, fifteen  or thirty minutes. These measurements are collected at a centralised location. The measurements are stacked together to form a data matrix, $\Zbf$, as follows:
\begin{equation}
\Zbf = \left[ \begin{array}{c}z_{ij} \end{array} \right]_{(n \times N)}
\end{equation}
where $z_{ij}$, henceforth denoted as $z_i^m(j)$, is the energy measurement corresponding to the $i^{th}$ node in the $j^{th}$ time interval, $n$ is the number of nodes in the network and $N$ is the number of measurements captured per node. 

\subsection{Energy Conservation} \label{Conservation}
\begin{figure}[h]
\centering
\psfrag{a}[c][][2]{ \tiny \hspace{-1mm}a}
\psfrag{b}[c][][2]{\tiny b}
\psfrag{c}[c][][2]{\tiny \hspace{1mm}c}
\psfrag{d}[c][][2]{\tiny d}
\psfrag{e}[c][][2]{\tiny e}
\psfrag{f}[c][][2]{\tiny f}
\psfrag{g}[c][][2]{\tiny g}
\psfrag{1}[c][][2]{\tiny 1}
\psfrag{2}[c][][2]{\tiny 2}
\psfrag{3}[c][][2]{\tiny 3}
\psfrag{4}[c][][2]{\tiny 4}
\psfrag{5}[c][][2]{\tiny 5}
\psfrag{6}[c][][2]{\tiny 6}
\psfrag{7}[c][][2]{\tiny 7}
\psfrag{8}[c][][2]{\tiny 8}
\includegraphics[scale=0.2]{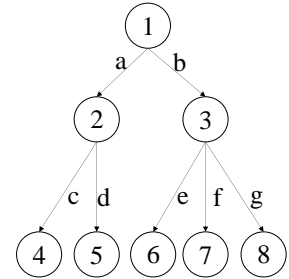}
\caption{A distribution network with  eight energy meters  connected through seven power lines }\label{EnergyCon} \vspace{-0.5cm}
\end{figure}

In this section, the concept of energy conservation will be illustrated using an example.
Consider a graph of a power network having eight energy meters (denoted as nodes 1, 2, $\ldots$, 8), connected through seven power lines (denoted as edges $a,\,b,\,\ldots,\,g$) as shown in Fig.~\ref{EnergyCon}. The incoming energy is captured by an energy meter at each of the nodes. 


The principle of conservation of energy implies that the sum of energies of incoming lines is equal to sum of energies of outgoing lines at any node. Assuming noise-free readings, the meter readings at the nodes $1, 2, \ldots, 8$ in Fig.~\ref{EnergyCon} can be related via the following equations, by applying the principle of conservation of energy, for all times $j=1, \ldots,N$:
\begin{eqnarray}
z_1^t(j) &=& z_2^t(j) + z_3^t(j) \\
z_2^t(j) &=& z_4^t(j) + z_5^t(j) \\
z_3^t(j) &=& z_6^t(j) + z_7^t(j)+ z_8^t(j).
\end{eqnarray}
Note that the parent node (meter) reading is equal to the sum of its child nodes (meters) readings in the graph of a distributed network due to the energy conservation. This principle leads a set of linear equations between the nodal readings described by:
\begin{equation}
z_k(j) = \sum\limits_{i} z_i(j), \; \forall \, k \in \mathcal{K}, \; i \in \mathcal{I}_k\label{Eq 4} 
\end{equation}
where $\mathcal{K}$ is the set of all parent nodes in the graph and $\mathcal{I}_k$ is the set of child nodes to the parent node $k$.

\subsection{Losses and Errors} \label{Errors}
In practice, we have to account for various sources of noise in the measurements to infer the underlying topology using data. Here, we account for the technical losses, random errors and clock synchronization errors in the smart meter readings, and consider them as sources of noise in the measurements.  These sources of noise and their contributions in the measurements are modelled next.

\subsubsection{Technical losses}
The technical losses include the constant losses such as iron losses, dielectric losses, and the variable losses such as copper losses. The latter vary with the load in the network and also depend on the length of the lines. These losses are modelled as errors in the readings as they affect the energy conservations between the nodal readings. Let $\boldsymbol{\lambda}$ be $n$--dimensional vector of technical losses for the $n$ lines in the   $j^{th}$ time interval. They can be modelled as  Gaussian with a non-zero mean, and heteroscedastic variance as follows: 
\begin{equation}
	\lambdabf(j) \sim \mathcal{N}(\mubf_\lambda, \Sigmabf_\lambda)
\end{equation}
where $\mubf_\lambda$ is the vector of means and $\Sigmabf_\lambda$ is called the error covariance matrix.
The mean captures the sum of the constant losses and the average of the variable losses while the variance captures the change in the variable losses. Since there are no correlations between losses in the different lines, $\Sigmabf_\lambda$ is a diagonal matrix with no covariance elements.

\subsubsection{Random errors in meter readings}
The latest ANSI standard for electricity meters stipulates that electricity meters must be of 0.2 or 0.5 accuracy class \cite{ANSI2010}. This indicates that  the meter reading can be in the range of $\pm\, 0.2\%$ and $\pm\,0.5\%$  of true values for 0.2 and 0.5 accuracy class meters, respectively. This error can also be  modelled to be Gaussian with each variable having a different error variance. The distribution of the error vector due to random errors $\epsilonbf(j)$ in the readings during the $j^{th}$ time interval is given by:  
\begin{equation}
\epsilonbf(j) \sim \mathcal{N}(\zeros,\Sigmabf_\epsilon) 
\end{equation}
where $\Sigmabf_\epsilon$ is a diagonal error--covariance matrix due to uncorrelated errors. 
\subsubsection{Clock synchronization errors (CSE)}
Though the clocks of all the meters are assumed to be synchronized, the synchronism may not be perfect leading to time intervals of energy measurements to be varying. The variation is generally in the order of milliseconds to few seconds. The error introduced by this variation is modelled to be a zero-mean Gaussian distribution as follows:
\begin{equation}
\deltabf(j) \sim \mathcal{N}(\zeros,\Sigmabf_\delta) 
\end{equation}
where $\Sigmabf_\delta$ is an $n$-dimensional diagonal matrix. 

The contribution of the total error  in the measurements is the summation of all errors from these sources. Hence, the measurements at the$j^{th}$ interval can be written as:
\begin{equation}
\zbf^m(j) = \zbf^t(j) + \lambdabf(j) + \epsilonbf(j) + \deltabf(j).
\end{equation}
It is assumed that the cross-correlation between the different components of error is negligible and we get,
\begin{equation}
\lambdabf(j) + \epsilonbf(j) + \deltabf(j) \sim \mathcal{N}(\bf{\mu}_\lambda,\Sigmabf_e),\,
\Sigmabf_e = \Sigmabf_\lambda + \Sigmabf_\epsilon + \Sigmabf_\delta. \label{ErrorCov} \end{equation}

\section{Phase and Topology Identification} \label{Solution}
Throughout this section, the following assumptions are made:
\begin{enumerate}
	\item There is no theft of electricity and there are no un-metered loads in the network.
	\item The topology of the underlying network remains unaltered while the $N$ measurements are captured.
\end{enumerate} \vspace{-0.2cm}
\subsection{Phase Identification}
In the phase identification problem, we can distinguish two kinds of nodes: (i) three nodes corresponding to the three phases of a transformer, (ii) consumer nodes. Since each  consumer node is connected to only one of the phases, the graph turns out to be a forest with three trees. Each tree has a parent node representing a phase and a number of child nodes representing consumers. Then, the problem of phase identification is to determine which child nodes are descendants of  which parent node.

Let us consider the forest shown in Fig.~\ref{Figure0}. In this forest, the  phase meters are parent nodes, and the consumer meters are child nodes. Then, the incidence matrix ($\Abf$) for this network is given by:
\renewcommand{\kbldelim}{(}
\renewcommand{\kbrdelim}{)}
\[ 
   \kbordermatrix{ 
    & a & b & c & d & e & f & g & h & i \\
    P1 & -1 & -1 & -1 & 0 & 0 & 0 & 0 & 0 & 0 \\
    P2 & 0 & 0 & 0 & -1 & -1 & -1 & 0 & 0 & 0 \\
    P3 & 0 & 0 & 0 & 0 & 0 & 0 & -1 & -1 & -1 \\
    C1 & 1 & 0 & 0 & 0 & 0 & 0 & 0 & 0 & 0 \\
    C2 & 0 & 1 & 0 & 0 & 0 & 0 & 0 & 0 & 0 \\
    C3 & 0 & 0 & 1 & 0 & 0 & 0 & 0 & 0 & 0 \\
    C4 & 0 & 0 & 0 & 1 & 0 & 0 & 0 & 0 & 0 \\
    C5 & 0 & 0 & 0 & 0 & 1 & 0 & 0 & 0 & 0 \\
    C6 & 0 & 0 & 0 & 0 & 0 & 1 & 0 & 0 & 0 \\
    C7 & 0 & 0 & 0 & 0 & 0 & 0 & 1 & 0 & 0 \\
    C8 & 0 & 0 & 0 & 0 & 0 & 0 & 0 & 1 & 0 \\
    C9 & 0 & 0 & 0 & 0 & 0 & 0 & 0 & 0 & 1 \\
  }
\]
The sub-matrix corresponding to the first three rows (related to the parent nodes) of the incidence matrix $\Abf$ provides the edge connectivity of the given network. Indeed, inferring this sub-matrix of $\Abf$ from measurements is sufficient to obtain the edge connectivity of the graph. This connectivity is unique according to Proposition \ref{Prop}. In general, the incidence matrix can be split as: 
\begin{equation}
\Abf = \left[ \begin{array}{c} \Abf_d \\ \Abf_i \end{array} \right]
\end{equation} 
where $\Abf_d$ are the rows corresponding to parent nodes and $\Abf_i$ are the rows corresponding to child nodes.


Due to the nature of phase and consumer measurements, the parent node variables can be taken as the dependent variables and the child node variables as the independent variables. With this notation, the regression matrix, $\Rbf$, given by PCA on measurements, relates the parent and child nodes in accordance with principle of energy conservation. It can be verified that the regression matrix $\Rbf$ which regresses  the dependent variables on the independent variables is in fact the matrix $\Abf_d$ with a negative sign. The uniqueness of $\Rbf$ makes it comparable element-wise to $-\Abf_d$ and hence, the connectivity of the underlying graph can be inferred from $\Rbf$.


PCA assumes the errors in measurements due to noise, to be \emph{i.i.d.}. Hence, to apply PCA on measurements, $\mubf_\lambda$, $\Sigmabf_\lambda$, $\Sigmabf_\epsilon$ and $\Sigmabf_\delta$ are to be estimated from data and $\Zbf$ needs to be pre-processed. 
$\mubf_\lambda$ has only three non-zero elements corresponding to the three phases because the technical losses do not reflect in the consumer readings. These elements are estimated from the mean of the total technical loss over all samples. Let $\mathcal{P}$ and $\mathcal{C}$ be the sets of rows of $\Zbf$ corresponding to phase nodes and consumer nodes, respectively. The mean is calculated from the difference in the summation of phase readings and the summation of consumer readings, as follows:
\begin{equation}
\hat{\mu}_t = \frac{\sum\limits_{j=1}^N \bigg(\sum\limits_{k} z_k^m(j)-\sum\limits_{i} z_i^m(j)\bigg)}{N}, \; \forall \, k \in \mathcal{P}, \; i \in \mathcal{C}.
\end{equation}
The non-zero elements of $\mubf_\lambda$, corresponding to each phase, are estimated as fractions of $\hat{\mu}_t$ proportional to the mean of respective phase readings. The element of $\mubf_\lambda$ corresponding to the $k^{th}$ phase is estimated as:
\begin{equation}  \label{meanloss}
\hat{\mu}_\lambda(k) = \hat{\mu}_t \frac{\sum\limits_{j=1}^N z_k^m(j)}{\sum\limits_{k} \sum\limits_{j=1}^N z_k^m(j)}, \forall \, k \in \mathcal{P}. 
\end{equation}
The pre-processing step includes separation of $\mubf_\lambda$ from each of the samples to ensure zero mean noise, as follows:
\begin{equation} \label{meanseparation}
\tilde{\zbf}(j)= \zbf^m(j) - \mubf_\lambda, \; \forall \, j = 1,\,\ldots, N. \vspace{-0.25cm}
\end{equation}
$\Sigmabf_\lambda$ has only three variance elements corresponding to the three phases. They are estimated from the variance in the total technical loss in fractions of variances of respective phase readings. The variance in total technical loss (denoted by $l_t$), is estimated as:
\begin{equation} \label{lossVar}
\Var[l_t] =  \frac{\sum\limits_{j=1}^N \bigg(\sum\limits_{k} z_k^m(j)-\sum\limits_{i} z_i^m(j) - \hat\mu_t\bigg)^2}{N}, \, \forall \, k \in \mathcal{P}, \, i \in \mathcal{C}.
\end{equation}
The diagonal element of $\Sigmabf_\lambda$ corresponding to the $k^{th}$ phase is estimated as:
\begin{equation}
\hat{\Sigma}_\lambda(k) =\Var[l_t] \frac{\Var[z_k]}{\sum\limits_{k} \Var[z_k]}, \; \forall \, k \in \mathcal{P}, \vspace{-0.25cm}
\end{equation}
where $\Var[z_k]$ is the variance in $N$ readings corresponding to $k^{th}$ phase.
The diagonal elements of $\Sigmabf_\epsilon$ are estimated based on the accuracy class of the meters. Let $\alpha$ be the accuracy class of a meter and as a result, the random errors in all the meter readings lie within $\alpha$ percentage of the reading. As nearly all values of a Gaussian distribution lie within three times its standard deviation, we estimate $\alpha$ percentage of the mean of the readings as three times the standard deviation of random errors. The diagonal element of $\Sigmabf_\epsilon$ corresponding to $i^{th}$ variable is estimated as:
\begin{equation}
\hat{\Sigma}_\epsilon(i) = \bigg(\frac{\alpha \bar{z_i}}{3 \times 100}\bigg)^2 \, \forall i \in \mathcal{P} \cup \mathcal{C},
\end{equation}
where $\bar{z_i}$ is the mean of the $i^{th}$ variable.
The standard deviation in the error due to imperfect time synchronization is taken as the deviation in the reading caused by one second change in the time interval. For each variable, this deviation is estimated from the mean of its readings. Hence, the diagonal element of $\Sigmabf_\delta$ corresponding to the $i^{th}$ variable is estimated as:
\begin{equation} \label{SyncVar}
\hat{\Sigma}_\delta(i) =\bigg(\frac{\bar{z_i}}{60T}\bigg)^2 \, \forall i \in \mathcal{P} \cup \mathcal{C}, 
\end{equation}
where $T$ is the time interval of a reading in minutes.

Now, $\Sigmabf_e$ is calculated following Eq.~\eqref{ErrorCov}. Let $\tilde{\Zbf}$ be the data matrix after the pre-processing step of error mean separation. PCA is applied on $\tilde{\Zbf}$ as described in Section \ref{ModelID} to estimate $\Rbf$ and the phase connectivity is inferred. 

The algorithm for phase identification is given as follows:
\begin{algorithm}
\caption{Phase Identification}
\begin{algorithmic}[1]
\State Start with $\Zbf$
\State Estimate $\mubf_\lambda$ as per Eq.~\eqref{meanloss}
\State Subtract $\mubf_\lambda$ from all columns of $\Zbf$ as per Eq.~\eqref{meanseparation} to get $\tilde{\Zbf}$
\State Estimate $\Sigmabf_\lambda$, $\Sigmabf_\epsilon$ and $\Sigmabf_\delta$ as per Eqs.~\eqref{lossVar} to \eqref{SyncVar}
\State Calculate $\Sigmabf_e$ using Eq.~\eqref{ErrorCov}
\State Compute $\hat{\Cbf}$ by applying PCA on $\tilde{\Zbf}$ following Eqs.~\eqref{Cholesky} to \eqref{Constraints}
\State Calculate $\hat\Rbf$ as per Eq.~\eqref{regression}
\State Round off $\hat\Rbf$ to truncate deviations due to noise and numerical residues. In each column, the element closest to 1 is rounded to 1 and rest 0.
\State Infer phase connectivity from $\hat\Rbf$
\State End
\end{algorithmic}
\end{algorithm} \vspace{-0.4cm}

\subsection{Topology Identification} \label{TopoID}
\begin{figure}[h]
\centering
\psfrag{V4}[c][][1.75]{\begin{tabular}{@{}l@{}}
   Voltage Level 4 \\
   (Layer 4)
\end{tabular}}
\psfrag{V3}[c][][1.75]{\begin{tabular}{@{}l@{}}
   Voltage Level 3 \\
   (Layer 3)
\end{tabular}}
\psfrag{VII}[c][][1.75]{\begin{tabular}[c]{@{}l@{}}
   Voltage Level 2 \\
   (Layer 2)
\end{tabular}}
\psfrag{V}[c][][1.75]{\begin{tabular}{@{}l@{}}
   Voltage Level 1 \\
   (Layer 1)
\end{tabular}}
\includegraphics[scale=0.5,width=80mm]{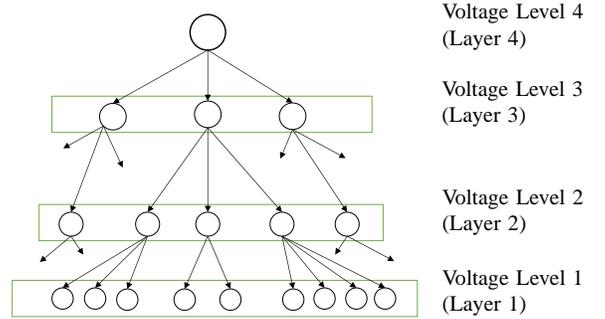} 
\caption{Layer-Wise Tree representation of Network Topology}
\label{LayerRep}
\end{figure}

The solution to the phase identification problem can be extended to the topology identification problem by  visualizing the tree structures in a layered manner. The nodes of the tree can be separated into layers with each layer having meters (nodes) operating at known voltage level, as indicated in Fig.~\ref{LayerRep}. Any set of two successive layers, when visualised separately, appears as a forest of directed trees. 

Now, the problem is reduced to finding connectivity of a forest of directed trees, which is similar to the phase identification problem. By inferring the connectivity between all possible successive layers, the complete network topology can be identified.

Let the layers be numbered from bottom to top as shown in Fig.~\ref{LayerRep}. Let $n_l$ be the number of layers and $N_l$ be the set of nodes present in layer $l$. Let $\zbf_i^T$ be the $i^{th}$ row of data matrix $\Zbf$. The following is the algorithm to topology identification:   
\begin{algorithm}
\caption{Topology Identification}
\begin{algorithmic}[1]
\State Start with $\Zbf$ and $l=1$.
\While {$l \leq n_l$} 
\State Let $\Zbf^* = {[\zbf_i^T]}^T \; \forall i \in N_{l+1} \cup N_l$
\State Estimate $\mubf_\lambda$ from $\Zbf^*$ as per Eq.~\eqref{meanloss}
\State Subtract $\mubf_\lambda$ from all columns of $\Zbf^*$ as per Eq.~\eqref{meanseparation} to get $\tilde{\Zbf}^*$
\State Estimate $\Sigmabf_\lambda$, $\Sigmabf_\epsilon$ and $\Sigmabf_\delta$ from $\Zbf^*$ as per Eqs.~\eqref{lossVar} to \eqref{SyncVar}
\State Calculate $\Sigmabf_e$ using Eq.~\eqref{ErrorCov}
\State Compute $\hat{\Cbf}$ by applying PCA on $\tilde{\Zbf}^*$ following Eqs.~\eqref{Cholesky} to \eqref{Constraints}
\State Calculate $\hat{\Rbf}$ as per Eq.~\eqref{regression}
\State Round off $\hat{\Rbf}$ to truncate deviations due to noise and numerical residues.
\State Let $\hat{\Rbf}_l = \hat{\Rbf}$ and $l=l+1$
\EndWhile
\State Infer the topology from $\hat{\Rbf}_1,...,\hat{\Rbf}_{L-1}$.
\State End
\end{algorithmic}
\end{algorithm}

\section{Simulation Results} \label{Simulation}
The proposed algorithms are demonstrated through simulations on noisy data sets. The simulations are conducted on MATLAB$^{\textregistered}$ 2014a.   
\subsection{Phase Identification} \label{PhaseIDSim}
The network is built using random number generators in MATLAB$^{\textregistered}$, as follows:
\begin{enumerate}
\item The number of consumers connected per phase are chosen randomly (uniformly) between 75 and 100.
\item The $N$ readings for each of the consumer meters are sampled from one of the three uniform distributions, with ranges $(0-100),(0-300)$ and $(0-500)$, to account for consumers with different ranges of loads.
\item Now, the $N$ readings for each of the three phase meters are determined by summation of the meter readings of consumers connected to them, respectively.
\item The relative distances of the consumers from the transformer are assigned randomly, from a set of numbers. The product of these distances with respective consumer readings is taken and scaled to the range $(5-10)$. As the technical losses depend on consumer loads and their distances from the transformer, these scaled products are taken as the percentages over the consumer readings to calculate losses. The losses are added to the phase readings appropriately.
\item The random errors are introduced by assuming 0.5 accuracy class meters. 
\item To account for synchronization errors, 15 minute time interval is assumed and Gaussian error is added, with standard deviation equal to the deviation in reading caused by one second change in the interval.
\end{enumerate}

\begin{figure}[h]
\centering
\psfrag{n}[c]{$n$}
\psfrag{t}[c]{$t$ in seconds}
\psfrag{N1}[l][][0.75]{$N = n_i$}
\psfrag{N2}[l][][0.75]{$N = 2n_i$}
\psfrag{N3}[l][][0.75]{$N = 3n_i$}
\psfrag{N4}[l][][0.75]{$N = 4n_i$}
\psfrag{N5}[l][][0.75]{$N = 5n_i$}
\includegraphics[width=70mm,scale=0.5,]{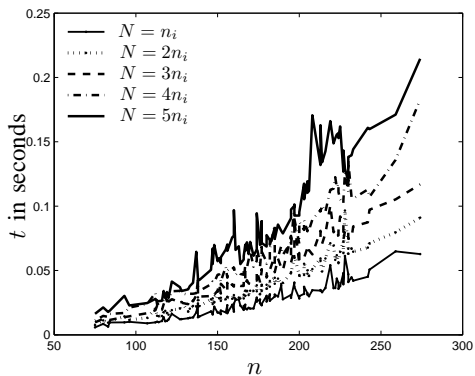}
\caption{No. of nodes Vs Simulation time}\label{Figure3}
\end{figure}

The algorithm is then applied to  hundred data sets, with different values of $N$ (multiples of $n_i$), and the time taken to arrive at the solution is noted in all the cases (Windows 10, Intel i5-4200U 1.64 Ghz processor, 6 GB RAM). The time taken to arrive at the solution against the number of nodes for different number of readings is plotted as shown in Fig.~\ref{Figure3}.

It can be observed from the Fig.~\ref{Figure3} that the time taken for phase identification is in the order of milli-seconds while an alternate method in \cite{Arya11}, which uses same type of data,  presents time taken to be in the order tens of seconds. Assuming that the computational power used in both the cases to be of same order, our method clearly outperforms the method proposed in \cite{Arya11}, in terms of time.


To compare the success rate of our algorithm with that proposed in \cite{Arya11}, phase identification of 10 randomly generated networks with upto 200 consumer nodes, was performed using both the algorithms and the success rates are reported in Table~\ref{Table2}. It can be observed that our method performs better on this front as well.  

\begin{table}[h]
\centering
\caption{Results of comparative simulations}
\label{Table2}
\begin{tabular}{|c|c|c|}
\hline
\multirow{2}{*}{\begin{tabular}[c]{@{}c@{}}No. of \\ Samples ($N$)\end{tabular}} & \begin{tabular}[c]{@{}c@{}}PCA based\\ method\end{tabular} & \begin{tabular}[c]{@{}c@{}}MIP based\\ method\end{tabular} \\ \cline{2-3} 
                                                                               & \multicolumn{2}{c|}{Success Rate (\%)}                                                                                  \\ \hline
$n_i$                                                                            & 0                                                          & 0                                                          \\ \hline
$2n_i$                                                                           & 100                                                        & 10                                                         \\ \hline
$3n_i$                                                                           & 100                                                        & 90                                                         \\ \hline
$4n_i$                                                                           & 100                                                        & 80                                                         \\ \hline
\end{tabular} \vspace{-0cm}
\end{table}
\subsection{Topology Identification}
The proposed algorithm for Topology Identification is tested by simulating data for the Bus 2 of Roy Billinton distribution test system \cite{Billinton91}, which has 2004 nodes as per our formulation.
The simulation is conducted as follows:
\begin{enumerate}
\item The $N$ readings for each of the consumer meters were sampled from a uniform distribution with mean and maximum equal to the average and peak loads of the consumers, as mentioned in \cite{Billinton91}.
\item The relative distances of the consumers from their source transformer and that of the transformers from their source feeders were randomly assigned. 
\item The transformer and feeder meter readings, at each of the $N$ time intervals, are then determined by appropriate summation of consumer meter readings.
\item The noise in the samples due to technical losses, random errors and time synchronization errors are added in a similar way, as described in Section \ref{PhaseIDSim}.
\end{enumerate}

The above simulation is repeated 10 times with different number of readings, $N$. The success rate and the average time taken for the algorithm to arrive at the solution, are shown in Table~\ref{Table 1}.
\begin{table}[h]
\centering
\caption{Topology identification: Simulation results}
\label{Table 1}
\begin{tabular}{|c|c|c|}
\hline
\begin{tabular}[c]{@{}c@{}}No. of \\ Samples ($N$)\end{tabular} & \begin{tabular}[c]{@{}c@{}}Success Rate\\ (\%)\end{tabular} & \multicolumn{1}{l|}{Average Time (sec)} \\ \hline
$n$                                                             & 10                                                         & 4.02                              \\ \hline
$2n$                                                            & 100                                                         & 7.47                              \\ \hline
$3n$                                                            & 100                                                         & 10.54                             \\ \hline
$4n$                                                            & 100                                                         & 13.91                             \\ \hline
$5n$                                                            & 100                                                         & 18.15                             \\ \hline
\end{tabular}
\end{table} \vspace{-0.4cm}

\section{Conclusion} \label{Conclusion}
In this paper, we proposed a novel data-driven approach for inferring the phase connectivity and network topology of an LV distribution network. The proposed approach uses  PCA and its graph theoretic interpretation to infer the topology from energy measurements. 
The proposed algorithms have been corroborated by simulation of random networks and also by simulating Roy Billinton distribution test system.

The proposed method infers the underlying topology accurately when sufficient data is available.  Further, the problem can be solved in the polynomial time, and hence, the solution can be transferred to practice in a straightforward manner. 

In the future, we  propose to use this technique for solving the problems of detecting changes in the topology, loss estimation, and detecting non-technical losses such as power theft.   We  also propose to extend this approach for inferring the underlying  network for missing data scenario.


%


\section*{Acknowledgment}
We would like to thank Prof S. Narasimhan  of IIT Madras for his valuable inputs.

\ifCLASSOPTIONcaptionsoff
  \newpage
\fi



%

\bibliographystyle{IEEEtran}
\bibliography{JournalBiB}

\begin{thebibliography}{10}
\providecommand{\url}[1]{#1}
\csname url@samestyle\endcsname
\providecommand{\newblock}{\relax}
\providecommand{\bibinfo}[2]{#2}
\providecommand{\BIBentrySTDinterwordspacing}{\spaceskip=0pt\relax}
\providecommand{\BIBentryALTinterwordstretchfactor}{4}
\providecommand{\BIBentryALTinterwordspacing}{\spaceskip=\fontdimen2\font plus
\BIBentryALTinterwordstretchfactor\fontdimen3\font minus
  \fontdimen4\font\relax}
\providecommand{\BIBforeignlanguage}[2]{{%
\expandafter\ifx\csname l@#1\endcsname\relax
\typeout{** WARNING: IEEEtran.bst: No hyphenation pattern has been}%
\typeout{** loaded for the language `#1'. Using the pattern for}%
\typeout{** the default language instead.}%
\else
\language=\csname l@#1\endcsname
\fi
#2}}
\providecommand{\BIBdecl}{\relax}
\BIBdecl

\bibitem{Schneider}
P.~John~Dirkman, ``Enhncing utility outage management system performance,''
  \emph{Schneider Electric White Paper}, 2014.

\bibitem{Jiyuan09}
J.~Fan, ``The evolution of distribution,'' \emph{IEEE Power and Energy
  Magazine}, vol.~7, pp. 63--68, 2009.

\bibitem{Kersting07}
W.~Kersting, \emph{Distribution system modeling and analysis}, 2nd~ed.\hskip
  1em plus 0.5em minus 0.4em\relax CRC Press, 2007.

\bibitem{Lueken12}
C.~Lueken, P.~M. Carvalho, and J.~Apt, ``Distribution grid reconfiguration
  reduces power losses and helps integrate renewables,'' \emph{Energy Policy},
  vol.~48, pp. 260--273, 2012.

\bibitem{Melo13}
F.~Melo, C.~Candido, C.~Fortunato, N.~Silva, F.~Campos, and P.~Reis,
  ``Distribution automation on lv and mv using distributed intelligence,'' in
  \emph{IEEE 22nd International Conference and Exhibition on Electricity
  Distribution}, 2013, pp. 1--4.

\bibitem{Cavaro15}
G.~Cavraro, ``Modeling, control and identification of a smart grid,'' Ph.D.
  thesis, University of Padova, 2015.

\bibitem{Dickson09}
K.~Dickson, ``Reduction of power losses using phase load balancing method in
  power networks,'' in \emph{World Congress on Engineering and Computer
  Science, San Francisco, USA}, 2009.

\bibitem{Das06}
D.~Das, ``A fuzzy multiobjective approach for network reconfiguration of
  distribution systems,'' \emph{IEEE Trasactions on Power Delivery}, vol.~21,
  pp. 202--209, 2006.

\bibitem{Jing12}
J.~Huang, V.~Gupta, and Y.-F. Huang, ``Electric grid state estimators for
  distribution systems with microgrids,'' in \emph{Annual Conference on
  Information Sciences and Systems (CISS), Princeton, USA}, 2012.

\bibitem{Chen11}
C.~S. Chen, T.~T. Ku, and C.~H. Lin, ``Design of phase identification system to
  support three-phase loading balance of distribution feeders,'' in
  \emph{Industrial and Commercial Power Systems Technical Conference
  (I$\&$CPS), Baltimore, USA}, 2011, pp. 1--8.

\bibitem{Zhiyu13}
S.~Zhiyu, M.~Jaksic, P.~Mattavelli, D.~Boroyevich, J.~Verhulst, and
  M.~Belkhayat, ``Three-phase ac system impedance measurement unit (imu) using
  chirp signal injection,'' in \emph{Applied Power Electronics Conference and
  Exposition (APEC), 2013 Twenty-Eighth Annual IEEE}, 2013.

\bibitem{Dilek02}
M.~Dilek, R.~P. Broadwater, and R.~Sequin, ``Phase prediction in distribution
  systems,'' \emph{IEEE Power Engineering Society Winter Meeting}, 2002.

\bibitem{Kezunovic06}
M.~Kezunovic, ``Monitoring of power system topology in real-time,'' in
  \emph{39th Hawaii International Conference on System Sciences}, 2006.

\bibitem{Arya11}
V.~Arya, D.~Seetharam, S.~Kalyanaraman, K.~Dontasn, C.~Pavlovski, S.~Hoy, and
  J.~R. Kalagnanam, ``Phase identification in smart grids,'' in \emph{IEEE
  International Conference on Smart Grid Communications, Brussels, Belgium},
  2011, pp. 1--6.

\bibitem{Arya13}
V.~Arya, T.~Jayram, S.~Pal, and S.~Kalyanaraman, ``Inferring connectivity model
  from meter measurements in distribution networks,'' in \emph{4th
  International Conference on Future Energy Systems}, 2013.

\bibitem{Pezeshki12}
H.~Pezeshki and H.~Wolfs, ``Consumer phase identification in a three phase
  unbalanced lv distribution network,'' \emph{IEEE PES Innovative Smart Grid
  Technologies, Europe}, 2012.

\bibitem{Tom13}
A.~Tom, ``Advanced metering for phase identification, transformer
  identification, and secondary modeling,'' \emph{IEEE Transactions on Smart
  Grid}, vol.~4, 2013.

\bibitem{Wen15}
M.~H. Wen, R.~Arghandeh, A.~von Meier, Poolla, and V.~O. Li, ``Phase
  identification in distribution networks with micro-synchrophasors,''
  \emph{IEEE Power and Energy Society General Meeting, Denver, CO}, 2015.

\bibitem{Wiel14}
S.~Wiel, R.~Bent, E.~Casleton, and E.~Lawrence, ``Identification of topology
  changes in power grids using phasor measurements,'' \emph{Applied Stochastic
  Models in Business and Industry}, vol.~30, no.~6, pp. 740--752, 2014.

\bibitem{Bolognani13}
S.~Bolognani, N.~Bof, D.~Michelotti, R.~Muraro, and L.~Schenato,
  ``Identification of power distribution network topology via voltage
  correlation analysis,'' in \emph{52nd IEEE Conference on Decision and
  Control, Florence, Italy}, 2013.

\bibitem{erseghe2013topology}
T.~Erseghe, S.~Tomasin, and A.~Vigato, ``Topology estimation for smart micro
  grids via powerline communications,'' \emph{IEEE Transactions on Signal
  Processing}, vol.~61, no.~13, pp. 3368--3377, 2013.

\bibitem{Aravind15}
A.~Rajeswaran and S.~Narasimhan, ``Network topology identification using {PCA}
  and its graph theoretic interpretations,'' in \emph{arXiv preprint
  arXiv:1506.00438}, 2015.

\bibitem{Jayadev16}
P.~S. Jayadev, A.~Rajeswaran, N.~P. Bhatt, and P.~Ramkrishna, ``A novel
  approach for phase identification in smart grids using graph theory and
  principal component analysis,'' in \emph{American Control Conference, Boston,
  USA}, 2016.

\bibitem{Jolliffe02}
I.~Jolliffe, \emph{Principal Component Analysis}, 2nd~ed.\hskip 1em plus 0.5em
  minus 0.4em\relax Springer-Verlay, New York, 2002.

\bibitem{Narasimhan08}
S.~Narasimhan and S.~Shah, ``Model identification and error covariance matrix
  estimation from noisy data using pca,'' \emph{Control Engineering Practice},
  vol.~16, pp. 146--155, 2008.

\bibitem{Narasimhan15}
S.~Narasimhan and N.~P. Bhatt, ``Deconstructing principal component analysis
  using a data reconciliation perspective,'' \emph{Computers $\&$ Chemical
  Engineering}, 2015.

\bibitem{Andrasfai91}
B.~Andrasfai, \emph{Graph Theory: Flows, Matrices}.\hskip 1em plus 0.5em minus
  0.4em\relax Akademiai Kiado, Budapest, 1991.

\bibitem{ANSI2010}
``Ansi c12.20-2010,'' \emph{American National Standard for Electricity Meters},
  pp. 1--11, 2010.

\bibitem{Billinton91}
R.~Allan, R.~Billinton, I.~Sjarief, L.~Goel, and K.~So, ``A realibility test
  system for educational purposes - basic distribution sytem data and
  results,'' \emph{IEEE Transactions on Power Systems}, vol.~6, no.~2, pp.
  813--820, 1991.

\end{thebibliography}

\vspace{-1.35cm}

\end{document}